\documentclass[journal]{IEEEtran}
\usepackage[dvips]{graphics}
\usepackage{graphicx}
\usepackage{amsmath}
\usepackage{cite}

\begin{document}
\title{Influence of Nuclear Spin Polarization on Quantum Wire Conductance}

\author{James A.\ Nesteroff, Yuriy V.\ Pershin and Vladimir
Privman \thanks{J.\ A.\ Nesteroff, Y.\ V.\ Pershin and V.\ Privman are
with Center for Quantum Device Technology,  Department of
Electrical and Computer Engineering, and Department of Physics,
Clarkson University, Potsdam, New York 13699-5721, USA; Electronic mail: nesteroff@clarkson.edu, pershin@clarkson.edu, privman@clarkson.edu.}
\thanks{This work was presented at the 2004 IEEE NTC Quantum Device Technology Workshop.}
}

\maketitle

\begin{abstract}
In this work, we study a possibility to measure the
transverse and longitudinal relaxation times of a collection of
polarized nuclear spins located in the region of a quantum wire
via its conductance. The interplay of an external in-plane
magnetic field, spin-orbit interaction, and the changing
field of the spin-polarized nuclei cause the conductance of the quantum
wire to evolve in time. We show that it is possible to extract the
transverse and longitudinal relaxation times of the spin-polarized
nuclei from the time dependence of the conductance.
\end{abstract}

\IEEEpeerreviewmaketitle

\section{Introduction}

In the present work we investigate the conductance of structures, which
fall into the realm of mesoscopic. This regime is characterized by length scales which
lie in between the microscopic and the more familiar macroscopic
world. Over the past decade and a half, there has been
considerable interest in these devices by both experimentalists
and theoreticians alike \cite{Kel,Ben}. Mesoscopic devices have
also found a place in the field of spintronics, which studies
novel electronic devices based on the use of electron spins
\cite{DatDas,DasSarma}.

The quantum wire (QW) is by far one of the simplest examples of
a mesoscopic device. The main principle is that it confines
electron transport to one dimension only. This can be accomplished
in a heterostructure, such as a AlGaAs/GaAs, which is used to
confine the electrons into a two dimensional electron gas (2DEG).
The application of split metallic gates on top of the 2DEG gives
rise to lateral confinement. Thus, the QW is formed in between two
reservoirs of electrons (or contacts). By applying a negative bias to
the gates, the electrons underneath are depleted, thus forming a
constriction in the 2DEG. When the width of this QW (constriction)
becomes short compared to the electron Fermi wavelength, a small number of
quantized modes appear. By changing the strength of the bias the
width of the QW, and ultimately the number of modes in
the channel, can be controlled. If the length of the quantum wire
is sufficiently short, the electrons  propagate through the device without
suffering collisions due to impurities, phonons, or other
electrons \cite{Datta}. This regime of transport is referred to as
\textit{ballistic} \cite{Kel,Ben,Datta}.

Some of the most interesting properties of QW appear in this
ballistic regime. For example, consider the situation when a bias
is applied across the contacts. Electrons are then allowed to flow
from one reservoir to the other through the QW. Since the
electrons do not suffer any scattering in the conductor one might
assume that there is zero resistance. Yet, this
conclusion is false, in fact there is a resistance which is
associated with the contacts themselves. Moreover, this resistance
can be expressed in terms of fundamental constants,
\begin{equation}
R = \frac{h}{2 e^2 M},
\end{equation}
\noindent where $M$ is the number of discrete modes in the
channel \cite{Datta,Exp1,Exp2}. For a single-mode wire this resistance is approximately
$12.9$k$\Omega$. Thus, in the limit of a
large number of modes in the channel this contact resistance
becomes negligible. However, if one were to vary the width of the
QW in such a way as to increase the number of modes one by one,
then the resistance (conductance) should decrease (increase) in a
step-like fashion.

The quantization of the conductance (or conductance plateaus) in
QWs was discovered in 1988 \cite{Exp1,Exp2}. Since then,
there has been a large number of studies, both
theoretical and experimental, of the properties of these
devices. In recent years, researchers have also started to
investigate the role which spin-orbit (SO) interaction, which
affects the electron spin, could play in the conductance of these
systems \cite{Moroz1,Moroz2,Moroz3,Streda,Pers,Nesteroff}. To the
best of our knowledge, the effect of SO interaction on QW
conductance was studied experimentally only in \cite{Yamada}.
However, the results reported in \cite{Yamada} do not allow to
clearly resolve the role of SO interaction in QW transport.

The SO interaction also plays a large part in the operation of
many recently proposed spintronic devices, reviewed, e.g., in
\cite{DasSarma}. One of the most popular and widely cited
proposals is the Datta-Das spin transistor \cite{DatDas}. Its
design is similar to the well known Field Effect Transistor (FET),
that is two contacts are connected by a channel (2DEG) with a gate
controlling the concentration of electrons in the channel. There
are, however, several important differences. The contacts are made
from magnetic materials The source contact causes electrons
entering the channel to become spin polarized. Secondly, the gate
controls the strength of the Rashba spin-orbit interaction
\cite{Rashba1,Rashba2}, as opposed to a regular FET in which the
gate only modulates the electron concentration. When an electron
enters the channel, the spin is rotated from its original
orientation by an amount proportional to the strength of the
Rashba coupling \cite{DatDas} and the length of the channel. This
phase shift can be, in principle, detected as a change in the
drain-source current, since the contact at the other end of the
channel acts as a spin filter \cite{DatDas}.

It has been shown that SO interaction can
modify the energy spectrum of the electrons in the channel of a
QW and, consequently, affect the conductance
\cite{Moroz1,Moroz2,Moroz3}. These studies were done under the
assumption of perfect transmission through the wire. Other
researchers \cite{Larsen,Mats,Bandy1,Bandy2} have focused on wires
which have transmission coefficients less than unity. In a recent
paper \cite{Pers}, we have shown that if an in-plane magnetic
field is applied to a QW in which the Rashba SO interaction is present then the
energy subbands inside the channel develop local extremal points
which are dependent on the magnetic field direction.

\begin{figure} [t]
\centering
\includegraphics[width = 8cm]{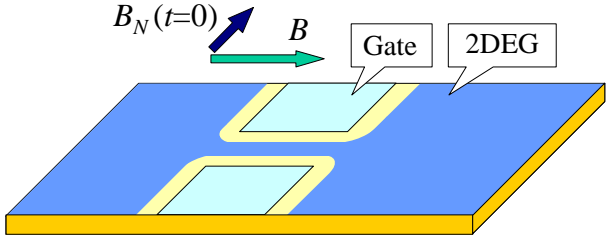}
\caption{Quantum wire with an applied magnetic
field (of magnitude $B$) in the $x$ direction, and an effective
nuclear hyperfine field (of magnitude $B_N$) initially pointing
in the $y$ direction.}
\label{scheme}
\end{figure}

Extending this idea further \cite{Nesteroff}, we have studied the
influence of a collection of spin-polarized nuclei in the region
of a QW on the conductance at low applied magnetic fields. A
possible realization of such a system is shown in Fig.\ \ref{scheme}. The nuclear spins are initially polarized in the direction
perpendicular to the QW and in the plane of the 2DEG. If a
magnetic field is applied along the QW, then the nuclear spins
will begin to precess around the magnetic field direction. The
influence of the polarized nuclear spins on the electron spins can be
described by an effective magnetic field, whose components will
change in time as the nuclear spins are reoriented. It was shown
that at low magnetic fields the precessing nuclear spin
polarization causes the extrema of the energy bands to change in time.
As a consequence, the conductance of the wire, at fixed bias,
shows an interesting time variation \cite{Nesteroff}. The aim of
this paper is to find the conductance in the limit of high
magnetic fields and long times as well as to summarize and extend
previously obtained results. We derive an expression for time
dependence of the conductance that can be used for extraction of
the transverse and longitudinal nuclear spin relaxation times from
experimental results.

\section{Overview of Physical Mechanisms}

In this section will give a brief overview of the theory of
transport in QW. This will be based on the well known
\textit{Landauer-B\"{u}ttiker} formalism. We will also include
an introduction to spin-orbit interaction concentrating on solid-state systems,
and briefly discuss nuclear spin polarization in semiconductors.

\subsection{The Landauer-B\"{u}ttiker formalism}

The transport properties of mesoscopic systems have been of great interest over the years \cite{Imry}.
Landauer was one of the first to relate the conductance
of a disordered one-dimensional system to the probability that
an electron can be transmitted through the system
\cite{Land1,Land2}. B\"{u}ttiker \textit{et
al.}\cite{Butt} generalized this idea to account for devices which
have an arbitrary number of leads and conducting channels in those
leads.

The electron energy
spectrum in the wire can be written as
$E_n(k)=E_n^{tr}+p^2/(2m^*)$, where $E_n^{tr}$ is the spectrum of
transverse subbands, $p$ is the momentum operator and $m^*$ is the
electron effective mass. Consider contribution to the current from
the electrons of a single subband. Starting from the relation for
current, $I = e\tilde{n}v$, where $e$ is the charge of the
electron, $\tilde{n}$ is the electron concentration, and $v$ is the
velocity of electrons in the channel, one can get the following
expression for the current through the QW,
\begin{equation}
I_n = \frac{2 e}{h} \int\limits_{E_n^{tr}}^{\infty} \left[
f(E,\mu_1) - f(E,\mu_2) \right] dE.\label{cur2}
\end{equation}
Here, $f(E,\mu_{1,2})$ is the Fermi distribution function
evaluated at the chemical potentials $\mu_{1,2}$ of the two
reservoirs that the QW connects. The prefactor
$2e/h$ in (\ref{cur2}) is a result of the cancelation of the
electron velocity and the one-dimensional density of states. Using
the relation, $G = I/V$, where $eV$, is the difference in the
chemical potentials of the two reservoirs, and taking into account
the contributions from all subbands, we can write the conductance
as \cite{Datta}
\begin{equation}
G = \frac{2 e^2}{h} \sum_{n=0} f(E_n^{tr},\mu). \label{lin4}
\end{equation}

In the zero temperature limit, (\ref{lin4}) becomes
\begin{equation}
G = \frac{2e^2}{h}M,\label{Gc}
\end{equation}
\noindent where, M is the number of modes in the channel (number
of $E_n^{tr}$ that is less than $\mu$).

The formulas (\ref{lin4}) and
(\ref{Gc}) were derived under the assumption that each subband has a
single extremal point and no spin dependence of the band structure.
A more generalized form of the Landauer-B\"{u}ttiker formula for the
conductance, which takes into account
energy bands with an arbitrary number of local extremal points and spin dependence,
is
\begin{equation}
G = \frac{e^2}{h} \sum_{n,s} \sum_{i} \beta_i^{n,s}
f(E^{n,s}_i,\mu),\label{Gspin}
\end{equation}
\noindent where the first sum is over each spin direction ($s$)
and each subband ($n$), the second sum is over subband extremal
points, $E^{n,s}_i$ denotes the extremal energy, and
$\beta_i^{n,s}$ is $+1$ for minima and $-1$ for maxima
\cite{Pers}.

\subsection{Spin-Orbit Interaction}

The spin-orbit
interaction is a contribution to
the electron Hamiltonian which describes the coupling of the spin and the
motional degrees of freedom \cite{Cohen}.
The standard form assumed for the SO term in the
Hamiltonian for a particle
moving in an external electric field is \cite{Cond}
\begin{equation}
H_{\rm SO} =
\frac{\hbar}{4m^2c^2}\left(\mathbf{\nabla} V \times \mathbf{p} \right)
\cdot \mathbf{\sigma}.\label{spino}
\end{equation}
Here $\mathbf{p}$ is the momentum operator, $m$ is the mass,
$V$ is the electric field potential energy, and
$\mathbf{\sigma}$ represents the vector of the Pauli spin matrices.

When working with crystal structures, there are two main sources
of SO coupling. The Dresselhaus SO interaction \cite{Dress} is
present in semiconductors lacking bulk inversion symmetry. When
restricted to a two-dimensional semiconductor nanostructure with
specific growth geometry ((100) growth direction \cite{Silsbee}),
this coupling is of the form
\begin{equation}
H_D = \beta(p_x\sigma_x - p_y\sigma_y),
\end{equation}
\noindent where $\beta$ is a constant and $p_{x,y}$ are the
momentum operators in the \textit{x} and \textit{y} directions.
Assuming that the confining potential in the region of the 2DEG
channel is linear, $V(x) = Cx$ where C is a constant, the Rashba
contribution to the SO interaction
\cite{Rashba1,Rashba2} is obtained, in the form
\begin{equation}
H_R = \alpha(p_y\sigma_x - p_x\sigma_y) \label{Rash}
\end{equation}
\noindent where $\alpha$ is known as the Rashba constant. Both the
Dresselhaus and Rashba SO couplings are responsible for lifting
the two fold spin degeneracy of the electrons. We limit our
consideration to the systems with only the Rashba interaction,
because even in zinc-blende semiconductors it is possible to
suppress the Dresselhaus coupling by the appropriate
heterostructure growth protocols \cite{Ohno}. Generally,
incorporation of the Dresselhaus interaction into our calculations
is straightforward.

\subsection{Nuclear Spin Polarization}

Elemental isotopes with non-zero nuclear spin are present in all
semiconductor materials. Although the nuclear spins are usually
disregarded in studies of spin-polarized transport \cite{Sayka1},
it is known that in some cases nuclear spin polarization could
have a significant effect on system properties. When nuclear spins
in a semiconductor are polarized, the electron spins feel an
effective hyperfine field, which lifts the spin degeneracy. For
example, for naturally abundant isotopes in GaAs this field
reaches $B_N=5.3\,$T in the limit when all nuclear spins are
completely polarized \cite{Paget}. The evolution of the nuclear
spin polarization can be described phenomenologically by the Bloch
equations \cite{Abrag},
\begin{equation}
\frac{d {\mathbf B_N^i}}{dt} = \gamma_i {\mathbf B_N^i} \times
{\mathbf B} - \frac{B_{N,y}^i\hat{y} + B_{N,z}^i \hat{z}}{T^i_2} -
\frac{B_{N,x}^i - B^i_0}{T^i_1}\hat{x}, \label{BlochEq}
\end{equation}
\noindent where the index $i=1,\ldots,p$ denotes different types
of nuclear spins, $\gamma_i$ denote the gyromagnetic ratios,
$\mathbf{B}$ is the external magnetic field, $B^i_0$ gives the
equilibrium values for the effective magnetic fields of the
nuclear spins, and $T^i_{1,2}$ are the longitudinal and transverse
spin relaxation times, respectively. The total magnetic field due
to the polarized nuclear spins is defined as ${\mathbf B_N}=\sum
_{i=1}^p {\mathbf B^i_N}$. The equilibrium (thermal) value of the
effective magnetic fields $B^i_0$ is rather small. Generally, the
following condition is fulfilled: $T_1\gg T_2$ \cite{Abrag}.

Assuming that only the nuclear spin isotope with $i=1$ has
non-zero polarization component in the $y$-direction at $t=0$, we can
easily solve the Bloch equations (\ref{BlochEq}) to obtain the
time dependence of the effective magnetic field of the
spin-polarized nuclei,
\begin{eqnarray}
B_{N,x}(t)  & =  & \sum _{i=1}^p (B^i_{N,x}(t=0)-B^i_0)
e^{-t/T^i_1}+B^i_0 ,\label{BNx}
\\ B_{N,y}(t)& = & B^1_{N,y}(t=0)e^{-t/T^1_2}\cos(\gamma_1 B t)  ,\label{BNy}\\
B_{N,z}(t) & = & - B^1_{N,y}(t=0)e^{-t/T^1_2}\sin(\gamma_1  B t) .
 \label{BNz}
\end{eqnarray}
\noindent Here $B^1_{N,y}(t=0)$ and $B^{1,\dots,p}_{N,x}(t=0)$ are
the initial values of the effective magnetic fields. In what
follows we will denote $ \gamma\equiv \gamma_1$. Experimentally,
the manipulation of nuclear spins can be accomplished by many
different techniques including optical pumping \cite{Paget,pump2}
and nuclear spin polarization by spin-polarized current
\cite{Kane1,Wald}.

\section{Effect of a Magnetic Fields and Spin-Orbit Interaction on
Quantized Conductance}

Let us consider the energy spectrum of a QW in which,
due to an asymmetric confinement potential perpendicular to the
2DEG, an electron experiences Rashba SO interaction \cite{Pers}.
Moreover, it is assumed that QW is subjected to an in-plane
magnetic field $\mathbf{B} = B_x\hat{x} + B_y\hat{y}$. We align
our coordinate system such that $\hat{x}$ points along the
direction of the electron transport and $\hat{y}$ is in the plane
of the 2DEG and transverse to the conductor; see
Fig.\ \ref{scheme}. Thus the one-particle
effective-mass Hamiltonian can be written as,
\begin{equation}
H = \frac{\hat{p}^2}{2m^{\ast}} + V(y) - \imath \alpha
\frac{\partial}{\partial x} \sigma_y + \Gamma \mathbf{\sigma}
\cdot \mathbf{B},\label{Ham}
\end{equation}
\noindent where, $\Gamma = g^{\ast}\mu_B /2$, $g^{\ast}$ is the
effective electron \textit{g} factor, and $\mu_B$ is the Bohr
Magneton. It should be emphasized that the in-plane magnetic field
does not enter into (\ref{Ham}) via the vector potential in this
approximation \cite{Pers}.

\begin{figure}
\centering
\includegraphics[width = 7cm]{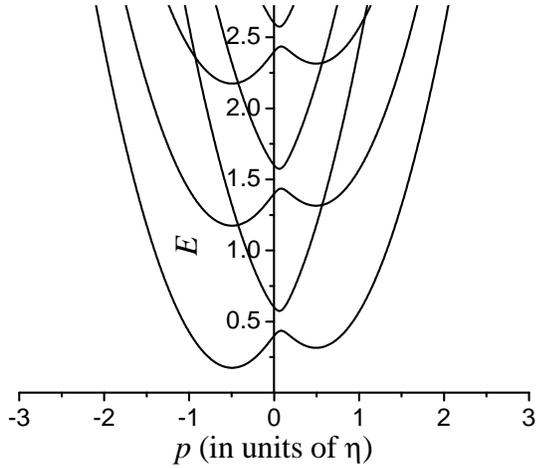}
\caption{Dispersion relation corresponding to $\theta = \pi/4$,
with energy defined in units of $\hbar\omega$, as a function of
the dimensionless momentum, $p/\eta$,
where $\eta = (2m^\ast \hbar \omega)^{1/2}$. The
parameters used in these plots were, $g^{\ast} \mu_B B / 2
\hbar\omega = 0.1$ and $\alpha[2m^\ast /(\hbar^3
\omega)]^{1/2} = 1$.}
\label{disp}
\end{figure}

The electron energy spectrum can be found by solving the {Schr\"odinger}
equation,
\begin{equation}
E = \frac{\hbar^2 k^2}{2m^{*}} + E_{n}^{tr} \pm \Gamma\sqrt{{B^2}
+ \frac{2 \alpha k B_y}{\Gamma} + \left( \frac{\alpha
k}{\Gamma}\right)^2}\label{Edisp}.
\end{equation}
\noindent Here, $B$ is the magnitude of the magnetic field, $B_y=
B\sin(\theta)$, $\theta$ is the angle of $\mathbf{B}$ measured
from the \textit{x}-axis and $E_{n}^{tr}$ are the eigenvalues of
the transverse modes. Assuming the parabolic confinement potential
in the $y$ direction, $V(y)$, we have
\begin{equation}
E_n^{tr}=\hbar\omega
(n+1/2).
\end{equation}
\noindent{}In Fig.\ \ref{disp}, we show the energy spectrum for
$\theta = \pi/4$.

It is interesting to analyze some of the features of the
dispersion relation (\ref{Edisp}) and its dependence on the
direction of the magnetic field. Let us start from the limiting
case when the magnetic field is aligned with the \textit{x}-axis.
In this case $E$ in (\ref{Edisp}) depends only on $k^2$, which
results in a symmetric band structure. If the magnetic field is
sufficiently small, then the lower band exhibits three local extrema
(one maximum and two minima). It is easy to note that the energy
splitting at $k=0$ is equal to the Zeeman splitting energy. As the
magnitude of the momentum of the electron increases, the effective
Rashba field becomes stronger and causes the spin quantization
axis to reorient itself along some linear combination of the
magnetic and effective Rashba field directions. In the limit of $k
\rightarrow \infty$ the spin quantization axis is completely
aligned with the Rashba field direction, which is along the
\textit{y}-axis. It is important to keep in mind that non-zero
$\theta$ results in the asymmetry of the dispersion relation (Fig.\ \ref{disp}). Maximal asymmetry is observed for $\theta=\pi / 2$
\cite{Pers}.

\begin{figure}
\centering
\includegraphics[width = 8cm]{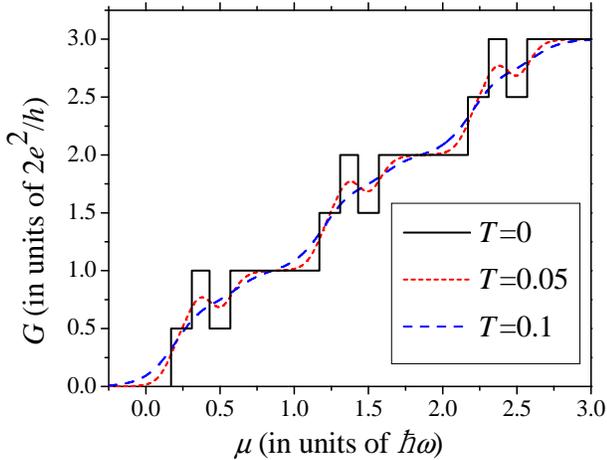}
\caption{Conductance as a function of the chemical potential,
$\mu$, for different temperatures $T$ (in units of $\hbar\omega /
k_B$), for $\theta = \pi/4$.} \label{calc}
\end{figure}

The conductance for $\theta = \pi/4$ is shown in Fig.\ \ref{calc}.
The features of the conduction
bands are reflected in the conductance, which differs drastically
from the conductance in which these effects are absent (see
\cite{Exp1,Exp2,Pep2}). If we consider the zero-temperature limit,
the rise of the chemical potential past each of the subband minima
results in the rise of conductance by $e^2/h$. When the chemical
potential passes a maximum, the conductance drops by $e^2/h$,
as given by (\ref{Gspin}). The effect of non-zero temperature
is a smearing of the sharp steps of conductance observed at $T=0$.

\section{Effect of Nuclear Spin Polarization on Conductance}

Having demonstrated that the conductance of a QW is significantly
effected by the presence of in-plane field and Rashba SO
interaction we now turn to an application of this finding. The main
goal of this section is to investigate how the nuclear spin
polarization influences the QW conductance, with particular
emphasis on nuclear spin relaxation time extraction from
time-dependence of the conductance. In what follows we assume that
the external magnetic field is applied in the $x$ direction, along
the wire, that is also the direction of the initial nuclear spin
polarization. After the nuclear spins were polarized, the nuclear
spins of a selected isotope are rotated in ($x,y$) plane from $x$
direction, via, for example, an NMR pulse \cite{Abrag} at $t=0$.
Influence of such non-equilibrium nuclear spin polarization on
quantum wire conductance is studied below.

In order to simplify our calculations, we assume that the
evolution of the nuclear spin polarization described by
(\ref{BNx})-(\ref{BNz}) occurs on a time scale that is much longer
than the time in which an electron traverses the QW. Thus, the
energy dispersion relations could be calculated utilizing an
adiabatic approximation. Then, the Hamiltonian of the system and
the electron energy spectrum are given by (\ref{Ham}) and
(\ref{Edisp}) respectively with the external magnetic field
$\mathbf{B}$ replaced by $\mathbf{B}_t=\mathbf{B}+\mathbf{B}_N$.

Below, we calculate the conductance of QW in two cases. We show
that in low magnetic fields, when the energy of the Zeeman
splitting is comparable with the relevant spin-orbit coupling
energy, the conductance of this system exhibits damped
oscillations on the time scale $T_2$ and a smooth evolution on the
time scale $T_1$. In the second case, of high magnetic fields, the
effect of the spin-orbit interaction can be neglected. Within this
approximation, the conductance evolution is smooth on both time
scales. We note that in order to observe the effect of the
nuclear spin polarization on the conductance, the gate voltage should
be properly selected: extremal subband points should be close to
the chemical potential of the QW.

\subsection{Low Magnetic Fields}

\begin{figure} [t]
\centering
    \includegraphics[width = 8cm]{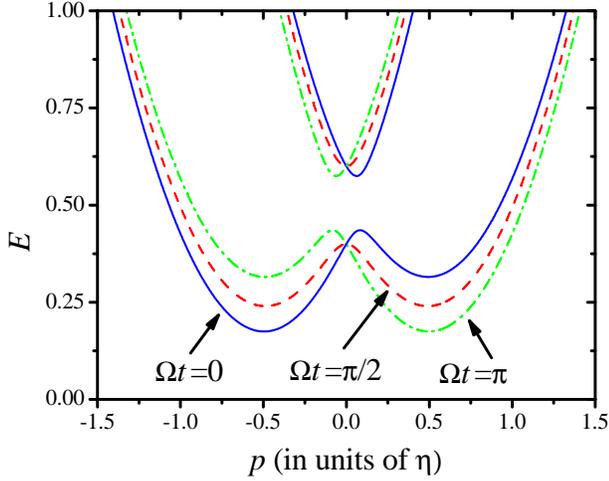}
    \caption{The lowest-energy sub-bands, in units of $\hbar\omega$, as functions of $p=\hbar k$,
for three different times. Here $\eta =
    (2m^{*}\hbar\omega)^{1/2}$ and $\Omega=\gamma B$. The two sets of curves correspond to spin
    $\pm$.}\label{b1}
\end{figure}

\begin{figure}[t]
\centering
    \includegraphics[width = 8cm]{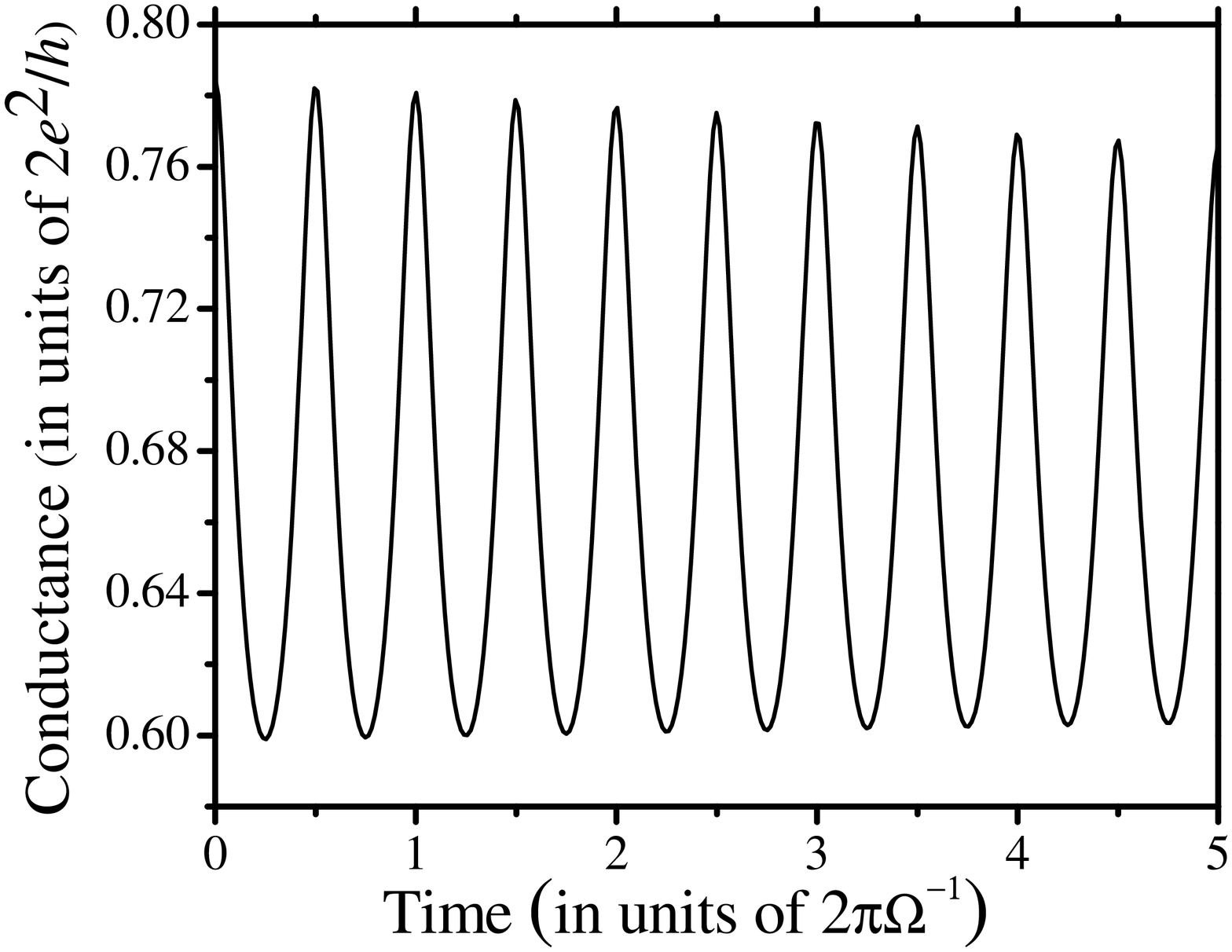}
    \caption{Time dependence of the conductance at a finite
    temperature and at the fixed value of $\mu = 0.5
    \hbar\omega$.}\label{b2}
\end{figure}

In this subsection we consider the effect of nuclear spin
polarization on QW conductance at short times (on the time scale
of $T_2$), characterized by non-zero transverse component of the
nuclear spin polarization, and at low magnetic fields. We have
already demonstrated in \cite{Nesteroff} that the
interplay of SO interaction with precessing nuclear spin
polarization results in time-dependent electron energy spectrum
and QW conductance. Fig.\ \ref{b1} shows the electron energy
spectrum at different moments of time. One can see that the time
dependence of the nuclear hyperfine field causes the energy spectrum
given by (\ref{Edisp}) to have oscillating extremal points in
time, as demonstrated in Fig.\ \ref{b1}. Only the local extrema
contribute to the conductance; see (\ref{Gspin}). Thus, the time
dependent conductance of this system will exhibit damped
oscillations, as shown in Fig.\ \ref{b2}. It is interesting to note
that there are two conductance oscillations per one period of
nuclear spin precession around the external magnetic field.

We can not calculate the extremal points of the
dispersion relation (\ref{Edisp}) in closed form; the results of
\cite{Nesteroff} were found numerically. It is instructive to get an
analytical expression for conductance, by considering the limit of
small transverse component of the nuclear field. Moreover, let
us consider the case when the chemical potential is such that the
main time-dependent contribution to the conductance is due to the
two local minima of the lowest subband; see Fig.\ \ref{b1}. This
condition is satisfied if the chemical potential $\mu$ is smaller
than the local maximum point of the first subband and the
temperature is sufficiently low. In this case, it is only
necessary to know the time dependence of the two local minima in
the dispersion relation.

Assuming that $B_{N,y}(t)$ is small, we can find approximate
positions of the minima of the dispersion relation (\ref{Edisp})
by setting $B_y\equiv B_{N,y}(t)=0$. The locations of the minima
of (\ref{Edisp}) are
\begin{equation}
k_m=\pm \sqrt{\frac{\alpha^2 m^{*2}}{\hbar^4}-\frac{B_t^2
\Gamma^2}{\alpha^2}}\label{min}.
\end{equation}
\noindent{}We substitute (\ref{min}) into (\ref{Edisp}) and expand the square
root in (\ref{Edisp}) in Taylor series. Keeping only the first-order terms we find,
\begin{equation}
E = \frac{\hbar^2 k^2_m}{2m^{*}} + E_{n}^{tr}
-\frac{\alpha^2m^*}{\hbar^2}-\frac{k_mB_{N,y}(t) \Gamma
\hbar^2}{\alpha m^*}.
 \label{app}
\end{equation}
\noindent{}We have compared the time dependence of the minima found
numerically vs.\ this approximation and found that for small
$B_{N,y}(t)$ the approximation is quite reasonable.

We are now in position to find the conductance of the system.
We are considering our system at low temperature and having
chemical potential below the local maximum in the first subband.
Therefore, to a good approximation the conductance can be expressed by
(\ref{Gspin}) as,
\begin{equation}
G = \frac{e^2}{h} \left[f(E_1) + f(E_2)\right]\label{app2},
\end{equation}
\noindent where, $E_1$ and $E_2$ are the energies of the two local
minima. If we note that (\ref{app}) can be written as $E=E_0
\pm \Delta E$, where $E_0 = \hbar^2 k^2_m/(2m^{*}) + E_{n}^{tr}
-\alpha^2m^*/\hbar^2$, $\Delta E =|k_m|B_{N,y}(t) \Gamma
\hbar^2/(\alpha m^*)$, and the $\pm$ correspond to the two
extremal points, then by substitution of this into
(\ref{app2}) we find,
\begin{equation}
G = \frac{2e^2}{h} \frac{1 + D \cosh(\beta(t))}{1 + D^2 + 2D
\cosh(\beta(t))}. \label{condapr}
\end{equation}
\noindent where $D = \exp \left[(E_0 - \mu)/(kT) \right]$,
$\beta(t) = \Delta E/(kT)$.

The evolution of nuclear spin polarization at longer times results
in smooth conductance variations on the time scale $T_1$. This
effect is similar to the smooth conductance variations discussed
in the next subsection.

\subsection{High Magnetic Fields}

\begin{figure}[bt]
\centering
\includegraphics[width=9cm]{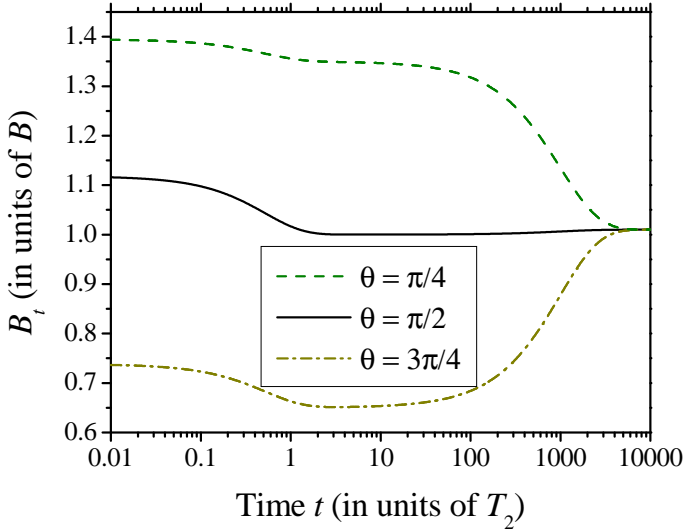}
\caption{Time dependence of $B_t$ calculated for the following set
of parameters: $B_0=0.01B$, $B_N(t=0)=0.5B$. Different curves
correspond to different initial directions of the nuclear spin
polarization.} \label{fen}
\end{figure}

Let us now consider the case when the applied magnetic field is
large enough or generally the cases when the effects of the Rashba SO
interaction can be neglected. By neglecting the third term in
(\ref{Ham}) and solving the {Schr\"odinger } equation via the adiabatic
approximation discussed earlier, we find that the eigenvalues
become
\begin{equation}
E_{n,\pm}(k) = \frac{\hbar^2 k^2}{2m^{*}} + E^{tr}_{n} \pm \Gamma
B_t(t), \label{spectr2}
\end{equation}
where $B_t$ is defined by
\begin{eqnarray}
\nonumber B_t^2(t) =
\left[B+B_0+(B_{N,x}(t=0)-B_0)e^{-t/T_1}\right]^2 +\\ B_{N,y}^2(t
= 0) e^{-2t/T_2},
\end{eqnarray}
and $B_0$ is the equilibrium value of the effective nuclear field.
Fig.\ \ref{fen} shows the time dependence of $B_t$ for different
values of $\theta$, the angle in the ($x,y$) plane between
the $x$ axis and the initial direction of the nuclear spin
polarization. Two different time scales are manifested in Fig.\ \ref{fen}. The short time decrease of $B_t$ is related to the
transverse nuclear spin relaxation, governed by the $T_2$ time, while
the long-time evolution is due to the longitudinal nuclear spin
relaxation, which occurs on the time scale $T_1$. From the
experimental point of view, the angle $\theta$ can be easily
controlled by varying the NMR pulse width.

The subbands given by (\ref{spectr2}) have only one minimum
point at $k=0$, for each spin. The energy associated with this
extremal point is simply $E_{n,\pm}(k=0)$. Let us assume that the
distance between the transverse subbands is much larger than the
temperature, $\hbar\omega \gg kT$, and that the chemical potential
$\mu$ is close to the minimum point of a subband. The
time-dependent contribution to the conductance, calculated by
substituting of the extremal points into (\ref{Gspin}), is given
by the same expression as (\ref{condapr}), but with the following
parameters: \ $\beta(t) = \Gamma B_t(t)/(kT)$ and $D = \exp
\left((E^{tr}_{n} - \mu)/(kT)\right)$.

\begin{figure}[tb]
\centering
\includegraphics[width=9cm]{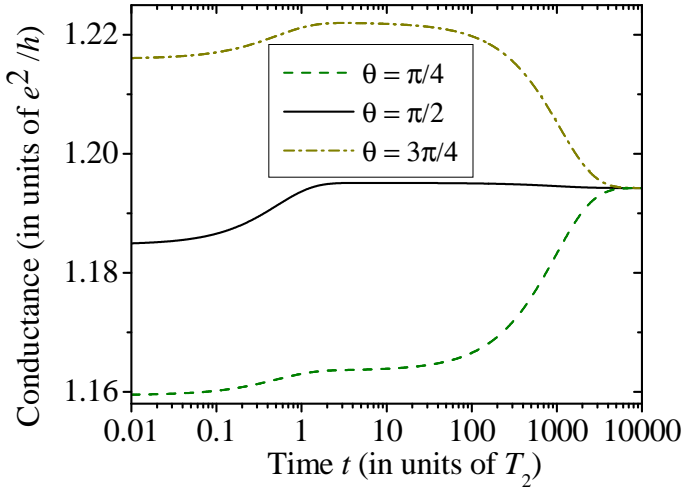}
\caption{Evolution of the conductance of the QW due to the
relaxation of the spin-polarized nuclei. Here
$(E_0^{tr}-\mu)/(kT)=0.5$, $\Gamma B/(kT)=1$, while the other parameters
are the same as in Fig.\ \ref{fen}.} \label{fcon}
\end{figure}
 
The time dependence of the conductance is defined by the time
dependence of the magnitude of the total field $B_t$, initial
magnitude of nuclear spin polarization $B_N$, its direction
described by the angle $\theta$, chemical potential, and
temperature. The time-dependent conductance of the system is shown
in Fig.\ (\ref{fcon}) for selected values of parameters and
different values of $\theta$. For these parameter values, for short
times the conductance $G$ initially increases with time. However the long time
behaviour is quite varying: conductance could be constant,
increasing or decreasing, depending on the initial direction of
the nuclear spin polarization. These interesting features open up
a way to experimentally separate out the contribution of a
selected nuclear isotope.

\section{Conclusions}

We have shown that via the time dependent conductance of a QW it
may be possible to extract the transverse and longitudinal
relaxation times of the spin-polarized nuclei in the channel of
the QW. When the effects of the Rashba SO interaction are
important, we find that the conductance exhibits damped
oscillations. The oscillations have a frequency that is
proportional to the gyromagnetic ratio of the nuclei and to the
applied magnetic field. The time scale of the damping depends
primarily on the time $T_2$. In the limit of a large applied
magnetic field, when the effects of the Rashba SO interaction can
be neglected, we have shown that the conductance oscillations
disappear. Our results suggest that it may be possible to extract
the $T_1$ and $T_2$ nuclear time scales from the experimental
data. The possibility to control the initial nuclear spin
polarization direction by an NMR pulse gives additional control
variable for interpretations of the experimental data. The
proposed technique can be useful for spin state readout in quantum
computing applications \cite{Shlim,PerQC,Priv2,Mani}.

\section*{Acknowledgment}

We acknowledge useful discussions with S.\ N.\ Shevchenko and I.\
D.\ Vagner. This research was supported by the National Security
Agency and Advanced Research and Development Activity under Army
Research Office contract DAAD-19-02-1-0035, and by the National
Science Foundation, grant DMR-0121146.

\bibliographystyle{IEEEtran}
\bibliography{mybib}

\end{document}